%% file: paper.tex
\documentclass{article}
\usepackage[T1]{fontenc}
\usepackage[utf8]{inputenc}
\usepackage{ismir}
\usepackage[bookmarks=false,hidelinks]{hyperref}
\usepackage{amsmath,cite,url,amssymb}
\usepackage{bm}
\usepackage{enumitem}
\usepackage{graphicx}
\usepackage{mathtools}
\usepackage{color}
\usepackage{booktabs,makecell}
\usepackage{tablefootnote}
\usepackage{placeins}
\usepackage{multirow}
\usepackage{xcolor}
\usepackage{tikz}
\usetikzlibrary{arrows.meta}

\usepackage[
  activate={true,nocompatibility},
  final,
  tracking=true,
  kerning=true,
  spacing=true,
  protrusion=true,
  expansion=true,
  factor=1200,
  stretch=30,
  shrink=30
]{microtype}
\microtypecontext{spacing=nonfrench}
\newcommand{\citeme}[1]{\textcolor{red}{[!]}}

\definecolor{claudegreen}{rgb}{0,0.5,0}

\title{Off-manifold robustness in synthesizer inversion with joint distribution flow matching}

\def\authorname{B. Hayes}
\oneauthor
  {Ben Hayes}
  {Sony Computer Science Laboratories Paris\\\texttt{ben.hayes@sony.com}}

\begin{document}

\maketitle

\begin{abstract}
Recent work on synthesizer inversion shows that generative models outperform deterministic approaches by explicitly modeling the ambiguity in mapping audio to parameters. 
Training such models, however, requires audio-parameter pairs, which are typically obtained by rendering sampled or preset parameters through the synthesizer itself.
This creates a train-test mismatch that can degrade performance on off-manifold real-world recordings, for which ground-truth parameter annotations do not exist.
To circumvent this obstacle, we propose to model the \textit{joint} distribution of audio and parameters with a multi-modal continuous normalizing flow using independent noise schedules for each modality.
This formulation allows us to train joint and conditional densities with paired synthesizer data, while unpaired real recordings can train the audio marginal alone, exposing the model to off-manifold signals without requiring parameter labels.
Further, because the model learns to map from audio to parameters at all noise levels, we find that partially noising the audio reference at inference improves real-audio reconstruction, consistent with reducing sensitivity to distribution-specific detail while preserving coarse structure.
Evaluating on {\sc Surge XT} and {\sc Dexed}, we find that modelling the joint distribution substantially improves both inversion of real-world and  in-domain audio.
\end{abstract}

\begin{figure}[t]
  \centering
  \includegraphics[width=\linewidth]{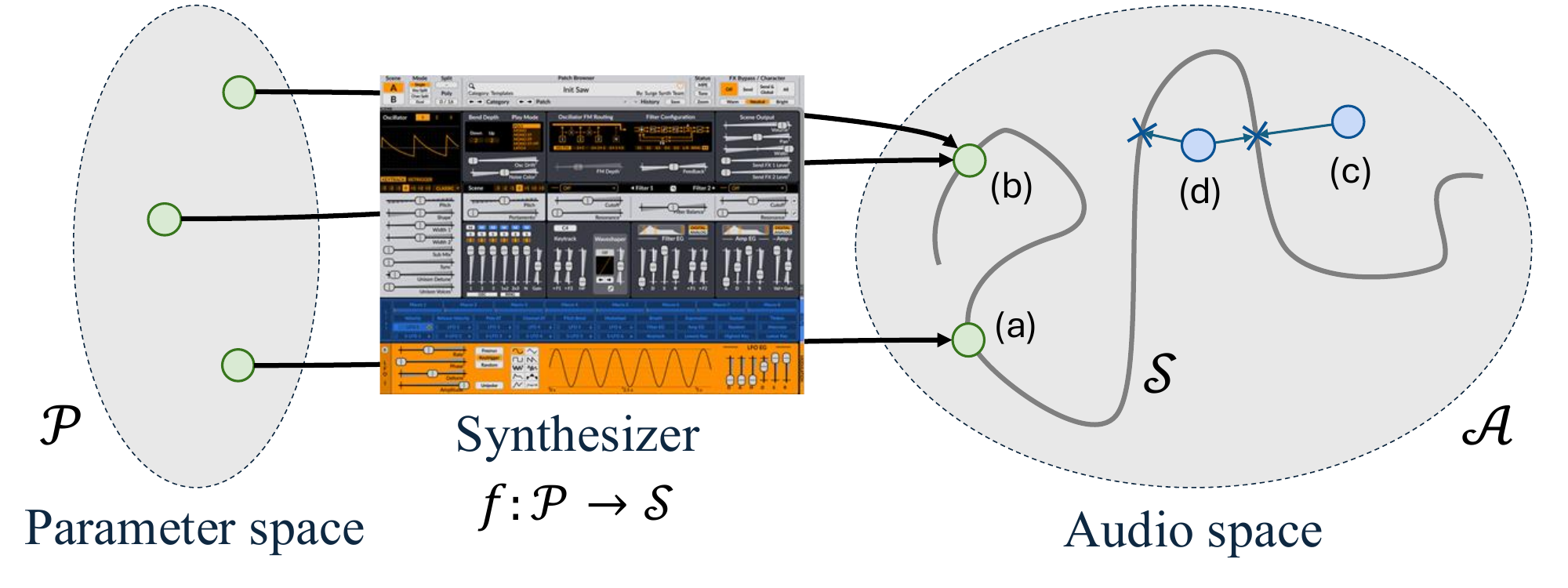}
  \caption{(a) An audio synthesizer $f$ maps from points in a parameter space $\mathcal{P}$ to points on a manifold $\mathcal{S}$ embedded in an ambient audio signal space $\mathcal{A}$. (b) Multiple points in $\mathcal{P}$ sometimes map to the same point on $\mathcal{S}$. (c) Off-manifold points in $\mathcal{A}$ can be optimally projected onto $\mathcal{S}$ under some distance metric. (d) Sometimes one off-manifold point admits multiple optimal projections, and sometimes multiple off-manifold points share an optimal projection on $\mathcal{S}$.}
  \label{fig:manifold}
\end{figure}

\section{Introduction}\label{sec:introduction}

A considerable body of work has sought to develop systems that automatically program audio synthesizers to recreate or approximate a given audio input~\cite{horner_machine_1993,barkan_inversynth_2019,barkan_inversynth_2023,esling_flow_2020,vaillant_improving_2021,vaillant_synthesizer_2023,masuda_synthesizer_2021,masuda_quality_2021,masuda_improving_2023,bruford_synthesizer_2024,hayes_audio_2025,shin_synthrl_2025,shier_synthesizer_2021,chen_sound2synth_2022,shier_spiegelib_2020}.
This task is intrinsically challenging: it is an ill-posed inverse problem~\cite{hayes_audio_2025} that must be solved under highly unfavourable conditions, as synthesizers are black boxes exhibiting strong nonlinearity, stochasticity, and complex parameter interactions. Recent work~\cite{hayes_audio_2025} has shown some of these difficulties can be mitigated by approaching the problem probabilistically, modelling the conditional distribution of parameters given audio while accounting for structural symmetries inherent to the synthesizer.

A gap remains, however, between these results and likely real-world use cases, in which a user presents the model with audio that was not produced by the synthesizer---i.e., audio that lies off the synthesizer's audio manifold (Fig.~\ref{fig:manifold}). Training such models requires audio paired with corresponding synthesizer parameters, and so training data is limited to audio produced by the synthesizer itself.
As we show empirically, this distribution mismatch is associated with degraded performance when the trained model is presented with off-manifold, real audio at inference time.

The obvious remedy would be to expose the model to real audio during training, but the natural mechanisms for doing so each carry significant costs. Differentiable digital signal processing~\cite{engel_ddsp_2020,hayes_review_2023,masuda_synthesizer_2021} permits direct supervision from real audio but sacrifices the benefits of generative training, and binds the method to specific differentiable implementations with their accompanying instability and optimisation challenges~\cite{hayes_review_2023,hayes_sinusoidal_2023,turian_im_2020}. 
Reinforcement learning approaches~\cite{shin_synthrl_2025} offer a complementary fine-tuning route, but synth-in-the-loop reward function computation renders them difficult to scale as a primary training objective, and requires substantial per-synthesizer design effort.

To sidestep these issues, we instead build on the generative synthesizer inversion framework~\cite{hayes_audio_2025} and repurpose joint multi-modal flow matching~\cite{li_omniflow_2025,bao_one_2023} as a mechanism for \emph{off-manifold robustness}.
Rather than treating audio as a fixed conditioning variable and learning only the conditional distribution of parameters given audio, we learn a single flow over the joint audio-parameter space with independent noise schedules for each modality.
While this framework was originally designed to unify multiple generative tasks under a single objective, we observe that its factorised noise schedule is also well suited to enhance the off-manifold robustness of a conditional generative model, for three reasons.
First, training on partially noised audio states constrains the velocity field away from the clean synthesized-audio manifold, reducing the extrapolation required when conditioning on real recordings.
Second, its independent noise schedules admit both the conditional parameter distribution and marginal audio distribution as special cases, allowing unlabeled real audio to be used alongside synthesizer audio as first-class training data.
Third, at inference, conditioning on a partially noised version of the target audio yields a noise-smoothed posterior that suppresses unreproducible recording-specific detail while preserving higher level structure.

We evaluate our approach on two widely-used software synthesizers, {\sc Surge XT} and {\sc Dexed}\footnote{{\sc Surge XT}: \url{https://surge-synthesizer.github.io/}; {\sc Dexed}: \url{https://asb2m10.github.io/dexed/}}, comparing against a prior generative inversion baseline~\cite{hayes_audio_2025} on both in-domain and out-of-domain audio. Our experiments demonstrate substantial improvements on real-world inputs while maintaining competitive in-domain performance. Audio examples are available at~\url{https://benhayes.net/synth-jdf/}.

\vspace{-1em}
\section{Background}\label{sec:background}

\begin{figure}[t]
  \centering
  \includegraphics[width=\linewidth]{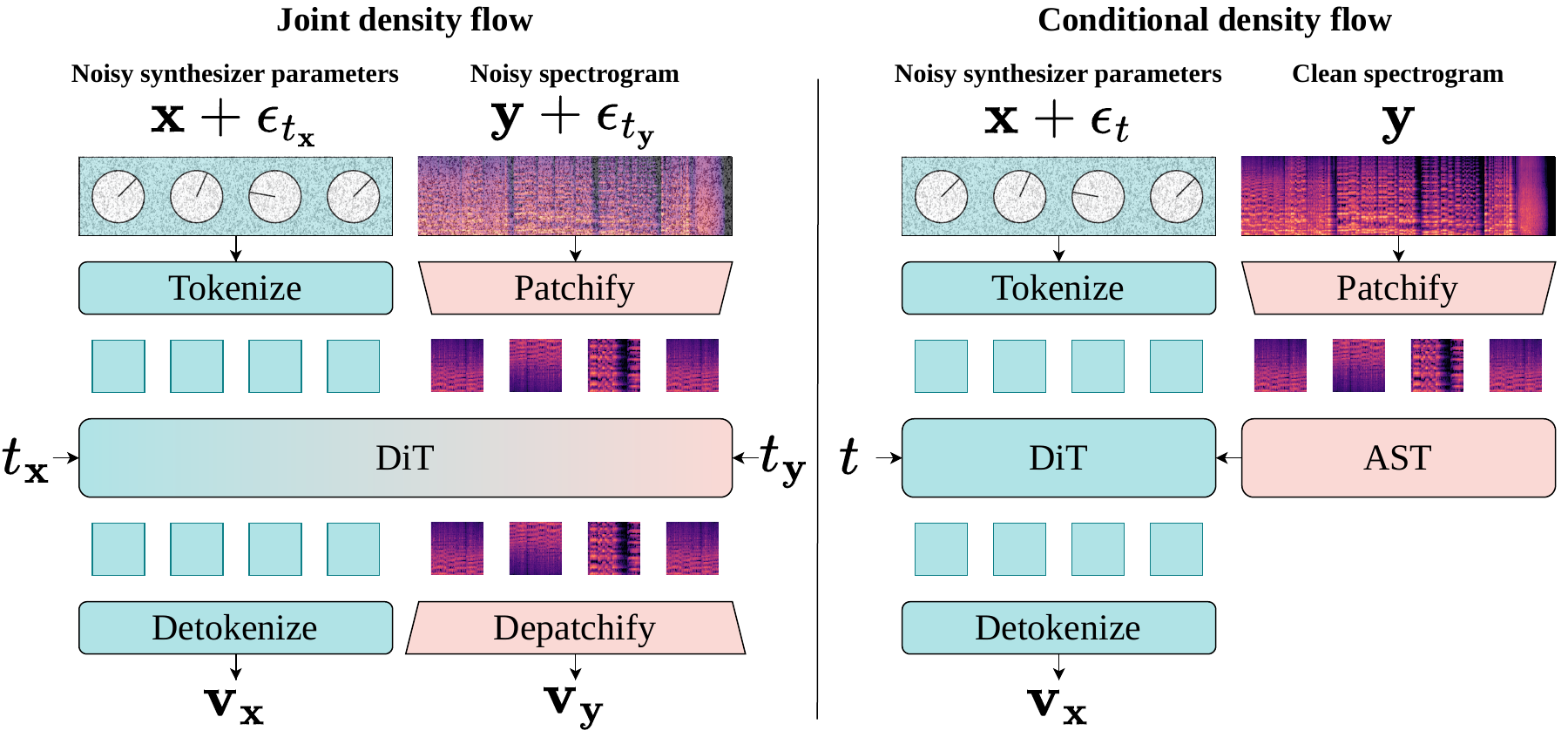}
  
  \caption{Left: Joint distribution flow matching for synthesizer inversion. Audio spectrogram patches and per-parameter tokens are concatenated and processed by a diffusion transformer (DiT) with per-modality time conditioning. Right: conditional distribution flow matching for synthesizer inversion.}
  \label{fig:architecture}
\end{figure}

\subsection{Synthesizer Inversion}\label{subsec:synth_inversion}

Let $\mathcal{P} \subset \mathbb{R}^k$ denote the space of synthesizer parameters and $\mathcal{A} \subset \mathbb{R}^n$ the space of audio signals. A synthesizer is the forward map $f: \mathcal{P} \to \mathcal{A}$. In synthesizer inversion, we seek to recover parameters $\mathbf{x} \in \mathcal{P}$ from a target signal $\mathbf{y} \in \mathcal{A}$.

In general, $f$ is non-injective: multiple parameter configurations can produce the same signal.
Deterministic regression methods~\cite{bruford_synthesizer_2024,yee-king_automatic_2018} therefore tend to collapse onto suboptimal solutions corresponding to averages over equivalent parameter sets.
Methods based on Differentiable Digital Signal Processing (DDSP)~\cite{masuda_synthesizer_2021,masuda_improving_2023,yang_white_2023} similarly rely on a deterministic mapping and cannot disambiguate equivalent or uncertain solutions, and require a differentiable synthesizer implementation.

Recently, the task has been approached generatively~\cite{hayes_audio_2025}, modelling the conditional distribution $p(\mathbf{x}\mid\mathbf{y})$ of parameters given audio. Generative approaches can place predictive mass on multiple plausible solutions, naturally accommodating the ill-posedness of the inverse problem while benefiting from the training stability of modern generative models. However, such models require training audio to be annotated with corresponding synthesizer parameters, and are thus exclusively trained on synthesized audio, which typically lies on a manifold $\mathcal{S}\subset\mathcal{A}$ (Fig.~\ref{fig:manifold}). This leads to degraded performance on real-world, off-manifold recordings.

Complementary work~\cite{shin_synthrl_2025} addresses domain gaps by fine-tuning a model with reinforcement learning on out-of-distribution audio for which parameter labels are unavailable.
However, this operates only on discrete parameter spaces and so we do not adopt it as a baseline.
We note, however, that our generative model could itself serve as a policy within such an RL framework.

\subsection{Flow Matching}\label{subsec:flow_matching}

Flow matching~\cite{liu_flow_2023,lipman_flow_2023} is a technique for training a continuous normalizing flow without relying on numerical integration during training.
Given data $X_1 \sim p_\mathrm{data}$ and independent noise $X_0 \sim p_0$, we define the stochastic interpolant:
\begin{equation}
    X_t = (1-t)X_0 + tX_1,\qquad t\in[0,1],
\end{equation}
with distribution $q_t$ at time $t$.
By construction, $q_0 = p_0$ and $q_1 = p_\mathrm{data}$. The \emph{marginal velocity field} is the conditional expectation
\begin{equation}\label{eq:marginal_velocity}
    u_t(\mathbf{x}) = \mathbb{E}\!\left[X_1 - X_0 \,\middle|\, X_t = \mathbf{x}\right],
\end{equation}
which satisfies the continuity equation $\partial_t q_t = -\nabla\!\cdot\!(u_t\, q_t)$. Transporting $p_0$ along $u_t$ from $t=0$ to $t=1$ therefore recovers $p_\mathrm{data}$.

A model $v_\theta$ is trained to regress $u_t$ via the flow-matching objective
\begin{equation}\label{eq:fm}
    \mathcal{L} = \mathbb{E}_{t, X_t}
    \left\| v_\theta(X_t, t) - u_t(X_t) \right\|^2.
\end{equation}
Direct minimisation is intractable since $u_t$ is unknown, but~\cite{liu_flow_2023,lipman_flow_2023} show that Eq.~\ref{eq:fm} shares its minimiser with the tractable conditional objective:
\begin{equation}\label{eq:cfm}
    \mathcal{L}_\mathrm{CFM} = \mathbb{E}_{t, X_0, X_1}
    \left\| v_\theta(X_t, t) - (X_1 - X_0) \right\|^2,
\end{equation}
whose targets are samples rather than conditional expectations. The minimiser of Eq.~\ref{eq:cfm} is, pointwise in $t$, the marginal velocity $u_t$. At inference, samples are generated by integrating $v_\theta$ from $t = 0$ to $t = 1$ with an ODE solver; by the continuity equation, this pushes $p_0$ forward to $p_\mathrm{data}$.

\subsection{Multi-modal Flow Matching}\label{subsec:multimodal_flow}

Bao et al.~\cite{bao_one_2023} observed that, given multi-modal data, diffusion models over marginal, conditional, and joint distributions can be unified by training a single model and assigning each modality its own independent noise level.
OmniFlow~\cite{li_omniflow_2025} extends this construction to flow matching: given $n$ modalities, each with its own time variable $t_i \in [0,1]$, different generation tasks correspond to different paths through the $n$-dimensional time cube $[0,1]^n$.
When a modality is unobserved, its time variable is set to $0$ (pure noise), and the model learns only from the remaining modalities.
This enables joint training on heterogeneous data in which different subsets of modalities are available for different examples.
The method we develop in Section~\ref{sec:method} builds directly on this framework, specialised to the two-modality setting of parameters and audio.


\vspace{-0.5em}
\section{Joint distribution flow matching for synthesizer inversion}\label{sec:method}
\input{figures/timesquare}



Let $X_1 \in \mathcal{P}$ and $Y_1 \in \mathcal{A}$ denote paired synthesizer parameters and audio, $(X_1, Y_1) \sim p_\mathrm{data}$, and let $X_0 \sim p_0^\mathbf{x}$ and $Y_0 \sim p_0^\mathbf{y}$ be noise samples, with $X_0$, $Y_0$, and $(X_1, Y_1)$ mutually independent. We assign each modality its own time variable, collected into the \textit{time square} $\mathbf{t} = (t_\mathbf{x}, t_\mathbf{y}) \in [0,1]^2$, and define the per-modality interpolants:
\begin{align}
\begin{split}
    X_{t_\mathbf{x}} &= (1-t_\mathbf{x})X_0 + t_\mathbf{x} X_1, \\
    Y_{t_\mathbf{y}} &= (1-t_\mathbf{y})Y_0 + t_\mathbf{y} Y_1.
\end{split}
\end{align}
Their joint distribution $q_\mathbf{t}$ is a family of distributions on $\mathcal{P} \times \mathcal{A}$ indexed by the time square. Its corners recover the data distribution $p_\mathrm{data}=q_{(1,1)}$, a pair of independent noise samples $p_0^\mathbf{x}\otimes p_0^\mathbf{y}=q_{(0,0)}$, and the clean-parameter and clean-audio marginals paired with the opposite modality's noise, i.e.\ at $\mathbf{t}=(1,0)$ and $\mathbf{t}=(0,1)$. Its edges recover conditional and marginal interpolants: the top edge $\{t_\mathbf{y}=1\}$ parameterises $p(X_{t_\mathbf{x}}\mid Y_1)$, the conditional interpolant from noise to clean parameters given clean audio; the left edge $\{t_\mathbf{x}=0\}$ factors as independent noise over $\mathcal{P}$ times the audio marginal interpolant $q_{t_\mathbf{y}}^\mathbf{y}$; and the right and bottom edges are their symmetric counterparts.

A model $v_\theta = (v_\theta^\mathbf{x}, v_\theta^\mathbf{y})$ predicting per-modality velocities is trained by the multi-modal conditional flow-matching objective of~\cite{li_omniflow_2025},
\begin{multline}
\label{eq:joint_loss}
    \mathcal{L}_\mathrm{joint} = \mathbb{E}_{\mathbf{t}\sim\mu}\!\left[
        \left\| v_\theta^\mathbf{x}(X_{t_\mathbf{x}}, Y_{t_\mathbf{y}}, \mathbf{t}) - (X_1 - X_0) \right\|^2 \right. \\ \left.
        + \left\| v_\theta^\mathbf{y}(X_{t_\mathbf{x}}, Y_{t_\mathbf{y}}, \mathbf{t}) - (Y_1 - Y_0) \right\|^2
    \right],
\end{multline}
where the expectation is also over $X_0, Y_0, X_1, Y_1$, and $\mu$ is a sampling distribution over the time square.
For unpaired real-audio examples, we restrict to the audio-marginal edge $t_\mathbf{x}=0$ and evaluate only the audio velocity term.
By the pointwise-minimiser property of the conditional FM objective~\cite{lipman_flow_2023} (Section~\ref{subsec:flow_matching}), $v_\theta$ approximates the marginal velocities of $q_\mathbf{t}$ at each $\mathbf{t} \in \mathrm{supp}(\mu)$, and is otherwise unconstrained.
Moreover, a sampling path $\gamma:[0,1]\to\mathrm{supp}(\mu)$ specifies which task is performed at inference: as $\gamma$ moves through the time square, only the velocity components corresponding to changing time coordinates are integrated~\cite{li_omniflow_2025}.

Since Eq.~\ref{eq:joint_loss} constrains $v_\theta$ only on $\mathrm{supp}(\mu)$, $\mu$ determines which distributions the model learns, as illustrated in Fig.~\ref{fig:flow-square}. 
Supporting $\mu$ on the full interior $[0,1]^2$ fits the joint distribution and its conditional and marginal paths.
Support only on the top edge recovers standard conditional flow-matching inversion, while support only on the left edge learns the audio marginal $p_\mathrm{data}^\mathbf{y}$ without parameter labels, allowing unpaired real recordings to be used directly. 

At inference, the sampling path $\gamma$ selects the task, as illustrated in Fig.~\ref{fig:flow-square}.
Integrating along the top edge with fixed reference audio $\mathbf{y}^*$ samples $p(\mathbf{x}\mid\mathbf{y}^*)$, the diagonal samples the joint distribution, and the left edge samples the audio marginal.
Interior horizontal paths $\{t_\mathbf{y}=\tau\}$ condition on partially noised audio, yielding a smoothed posterior that trades reference specificity against robustness to off-manifold detail.

\section{Experiments}\label{sec:experiments}

We evaluate our method on both \emph{in-domain} (synthesized) and \emph{off-manifold} (real-world) audio to assess its ability to generalize beyond the synthesizer manifold. Our setup is designed to isolate the effect of joint distribution flow matching and the inclusion of unpaired audio, while controlling for architecture and training budget.

\subsection{Data}
We evaluate on two software synthesizers spanning distinct synthesis paradigms.\footnote{We run full ablations on {\sc Surge XT} only, to make optimal use of available computational resources.}
{\sc Surge~XT} is a hybrid subtractive/wavetable synthesizer used as the
primary testbed in~\cite{hayes_audio_2025}.
{\sc Dexed} is an open-source clone of the Yamaha DX7, a six-operator
frequency modulation synthesizer.
FM synthesis~\cite{chowning_synthesis_1973} poses a challenge for parameter estimation algorithms~\cite{caspe_ddx7_2022,chen_sound2synth_2022}: small parameter perturbations can produce large, non-local timbral changes, and the operator routing (``algorithm'') reorganises the entire signal flow.
Prior work on {\sc Dexed} inversion has therefore restricted the problem by fixing or limiting the algorithm~\cite{vaillant_improving_2021,shin_synthrl_2025,chen_sound2synth_2022,shier_spiegelib_2020}.
We make no such restriction: our model jointly predicts the algorithm
alongside all continuous parameters, across all 32 routings.
This is a substantially harder setting, as the semantics of most parameters depend on the algorithm.

Following~\cite{hayes_audio_2025}, we generate paired audio--parameter examples by sampling parameters uniformly over their respective ranges and rendering audio through the synthesizer.
For {\sc Surge XT} we adopt a similar set of active parameters as in~\cite{hayes_audio_2025}, but disable audio effects.
For both synthesizers, we disable any modulator settings known to introduce non-determinism (e.g. Sample \& Hold).
Rather than fixing a dataset size, we render in-domain audio online during training.
Uniform parameter sampling produces a long tail of silent or near-silent outputs, which would otherwise dominate training; we therefore reject any
sample with broadband RMS below $-60$\,dBFS.
For validation and testing we fix a random seed and freeze a $10$k-example
set per synthesizer, ensuring determinism across model comparisons.

Continuous parameters are linearly mapped to $[-1, 1]$, and discrete
parameters (e.g., {\sc Dexed} algorithms, Surge oscillator types) are one-hot
encoded.
Audio is rendered at $44.1$\,kHz in stereo with a duration of $3.0$ seconds.
Inputs to the model are log-mel spectrograms with a $25$\,ms window and
$10$\,ms hop.
To avoid hand-tuned input normalisation, we estimate channel-wise mean and
variance over the first $8$k training spectrograms using Welford's online
algorithm and freeze the resulting statistics for the remainder of training.

\subsubsection{Off-manifold audio}\label{subsubsec:offmanifold_data}
Our central claim concerns inversion of audio that does not lie on the
synthesizer's manifold.
To stress-test this regime, we curate a dataset that approximates the
intended deployment setting --- short, isolated sounds of the kind a user
would plausibly hand to an inversion system.
We draw from two sources.
First, we filter Freesound~\cite{font_corbera_freesound_2013} for files flagged as containing a single
acoustic event by the Audio Commons single-event descriptor~\cite{pearce_timbral_2017},
discarding loops, ambiences, and multi-event recordings.
Second, we include a proprietary library of one-shot samples used in music
production, covering instruments, percussion, and sound-design material.
After deduplication, trimming/padding to $3.0$\,s, and resampling to
$44.1$\,kHz, we obtain $617{,}016$ training files together with held-out
validation and test sets of $10$k files each.
This data carries no parameter annotations and is consumed exclusively along
the audio-marginal edge $\{t_\mathbf{x}=0\}$ defined in
Section~\ref{sec:method}.

\subsection{Model variants}\label{subsubsec:variants}
We compare four models to isolate the effects of joint training, real-audio marginal supervision, and architectural unification.
Across the joint-density flow (JDF) variants, the only training-time difference is the time-sampling distribution $\mu$.
Exact task-mixture weights for $\mu$ are given in the supplementary material.

\begin{itemize}[leftmargin=*,itemsep=0pt,topsep=0pt]
    \item \textbf{JDF}: Our joint formulation, trained on the full time square $\mathrm{supp}(\mu)=[0,1]^2$.

    \item \textbf{JDF-A}: JDF with additional training on the \textbf{audio}-marginal edge $\{t_\mathbf{x}=0\}$ using both synthetic and unpaired real audio.

    \item \textbf{Unif-{\sc ParamOnly}}: Matches the unified JDF architecture, but trained only on the conditional inversion edge $\{t_\mathbf{y}=1\}$.

    \item \textbf{Conditional}: An encoder-based generative inversion baseline following~\cite{hayes_audio_2025}, where an audio spectrogram transformer (AST)~\cite{gong_ast_2021} audio encoder conditions a parameter (DiT)~\cite{peebles_scalable_2023} trained by conditional flow matching over $p(\mathbf{x}\mid\mathbf{y})$.
\end{itemize}

Fig.~\ref{fig:architecture} compares the unified JDF architecture to the encoder-conditioned baseline.

\subsubsection{Architecture and training}
All models are trained for 1.5M steps using a rectified flow parameterisation~\cite{liu_flow_2023} with linear interpolants.
Except where they are fixed to a constant value, time variables are sampled from a logit-normal schedule~\cite{esser_scaling_2024}.

Classifier-free guidance (CFG)~\cite{ho_classifier-free_2021} is supported in all variants.
For JDF and JDF-A, the parameter-marginal edge $\{t_\mathbf{y}=0\}$ provides a natural unconditional branch.
For Unif-{\sc ParamOnly} and Conditional, which are not trained on this edge, we use $10\%$ conditioning dropout by replacing audio conditioning with pure noise for the former, and a learnt CFG token for the latter.
We sweep guidance weight in Fig.~\ref{fig:guidance}.
Further training and architecture details are provided in the supplementary material.

\subsubsection{Inference}
All samples are drawn by integrating $v_\theta$ with an Euler solver for $20$ steps along the appropriate path in $[0,1]^2$.
For inversion, this is the top edge $\{t_\mathbf{y}=1\}$ with audio fixed at the (possibly partially noised) reference.
The conditioning noise level $\tau$ (Section~\ref{sec:method}) and the CFG weight are treated as inference-time hyperparameters and swept in Figs.~\ref{fig:noise_sweep}~and~\ref{fig:guidance}.
CFG is applied on the limited interval $[0.15, 0.95]$~\cite{kynkaanniemi_applying_2024}.

\subsection{Metrics}
We report the four reconstruction metrics used by Hayes et al~\cite{hayes_audio_2025} between each input audio signal and its resynthesis obtained by passing the inferred parameters back through the synthesizer.
These are the multi-scale spectral distance (MSS), warped-MFCC distance (wMFCC), spectral optimal-transport distance (SOT), and the cosine similarity between RMS-energy envelopes (RMS).
These are detailed further in the supplementary material.
Each metric is reported on two test conditions of $10$k examples each: an in-distribution split drawn from the synthetic data generator, and the off-manifold split of Section~\ref{subsubsec:offmanifold_data}.

\section{Results}\label{sec:results}
\input{tables/main}

\begin{figure}[t]
  \centering
  \includegraphics[width=\columnwidth]{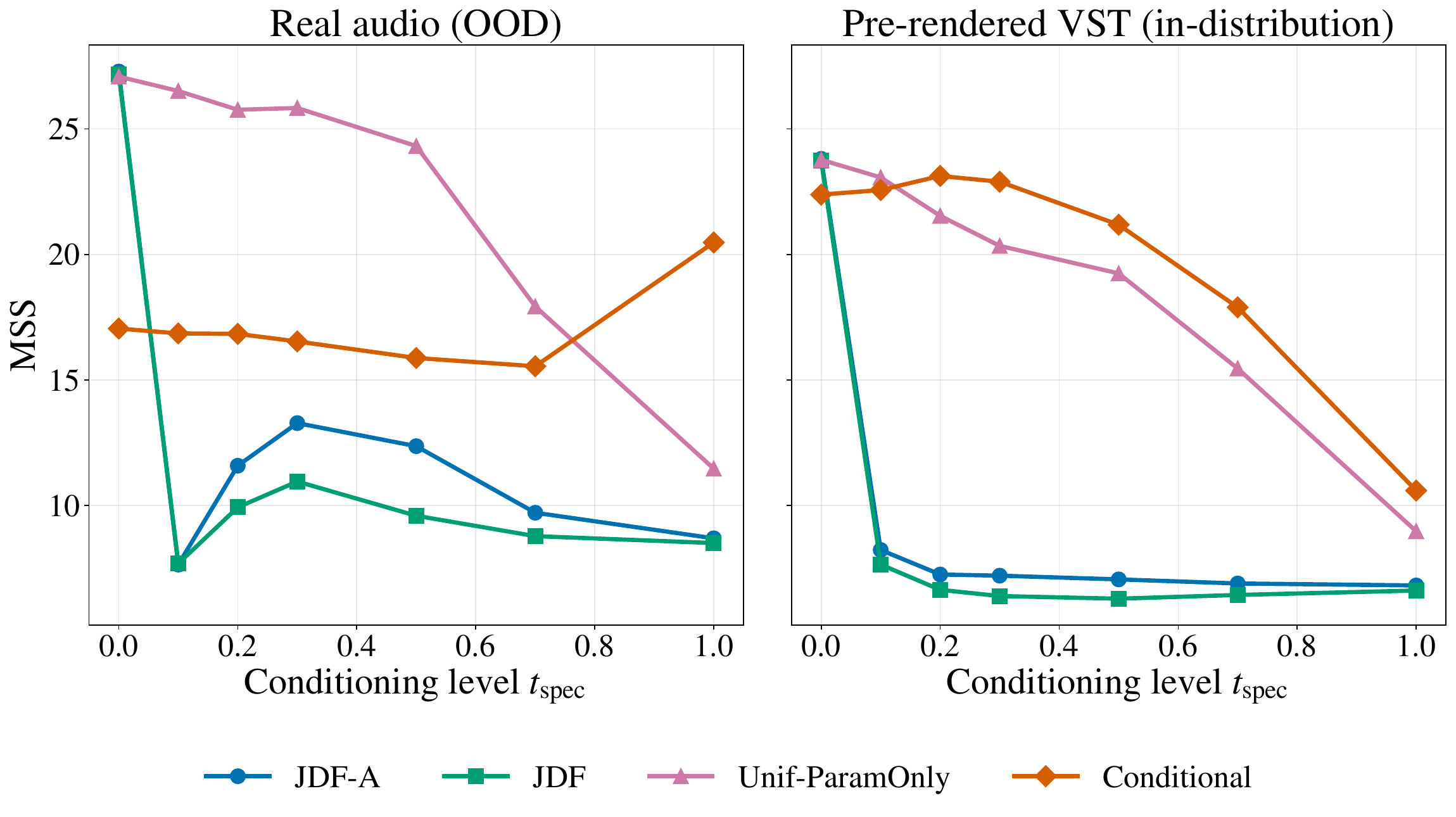}
  \caption{Effect of conditioning noise level on MSS.}
  \label{fig:noise_sweep}
\end{figure}

\begin{figure}[t]
  \centering
  \includegraphics[width=\columnwidth]{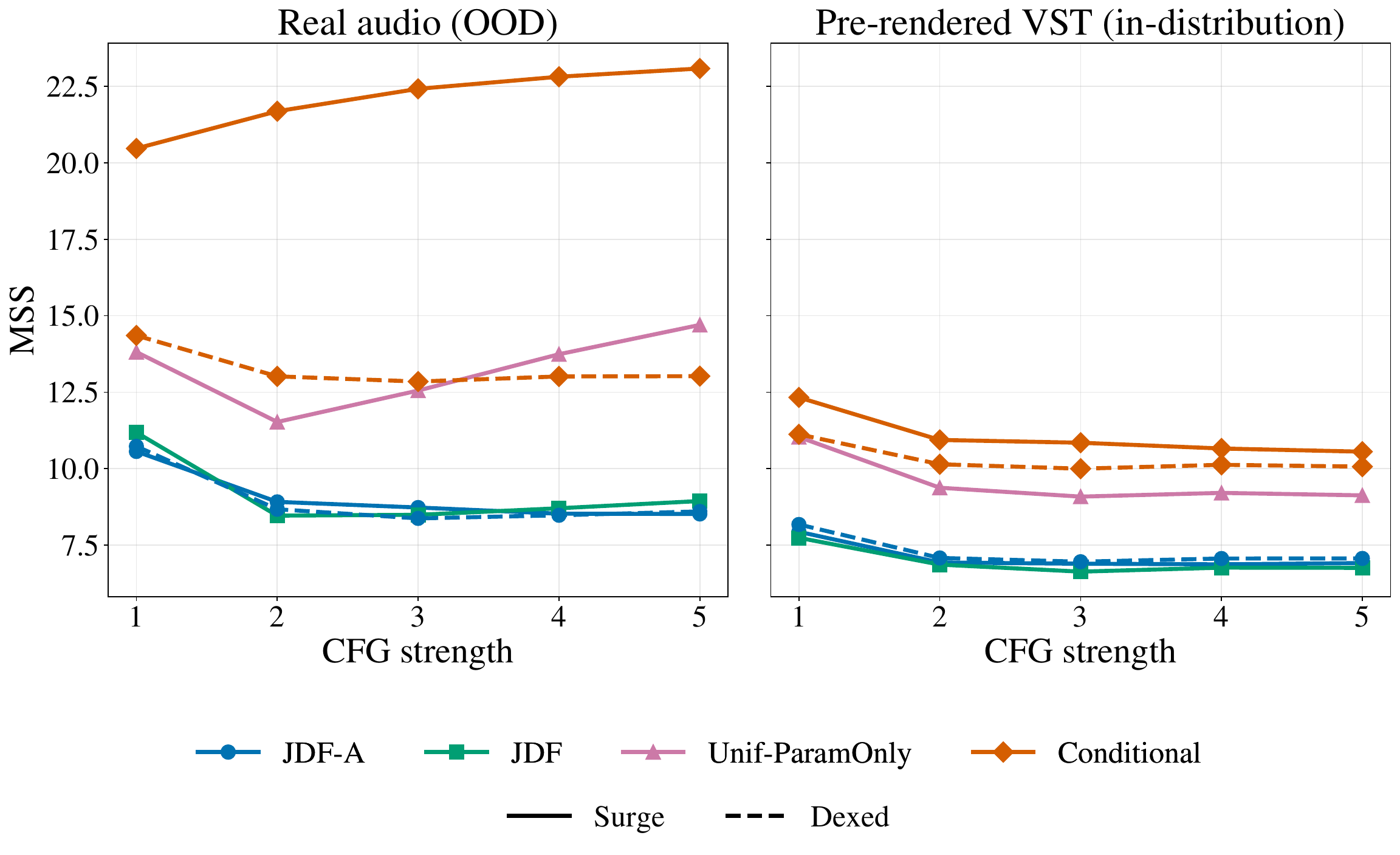}
  \caption{Effect of classifier-free guidance (CFG) weight. CFG is available for JDF and JDF-A models without conditioning dropout, as fully noised audio creates a natural unconditional branch.}
  \label{fig:guidance}
\end{figure}

\begin{figure}[t]
  \centering
  \includegraphics[width=\columnwidth]{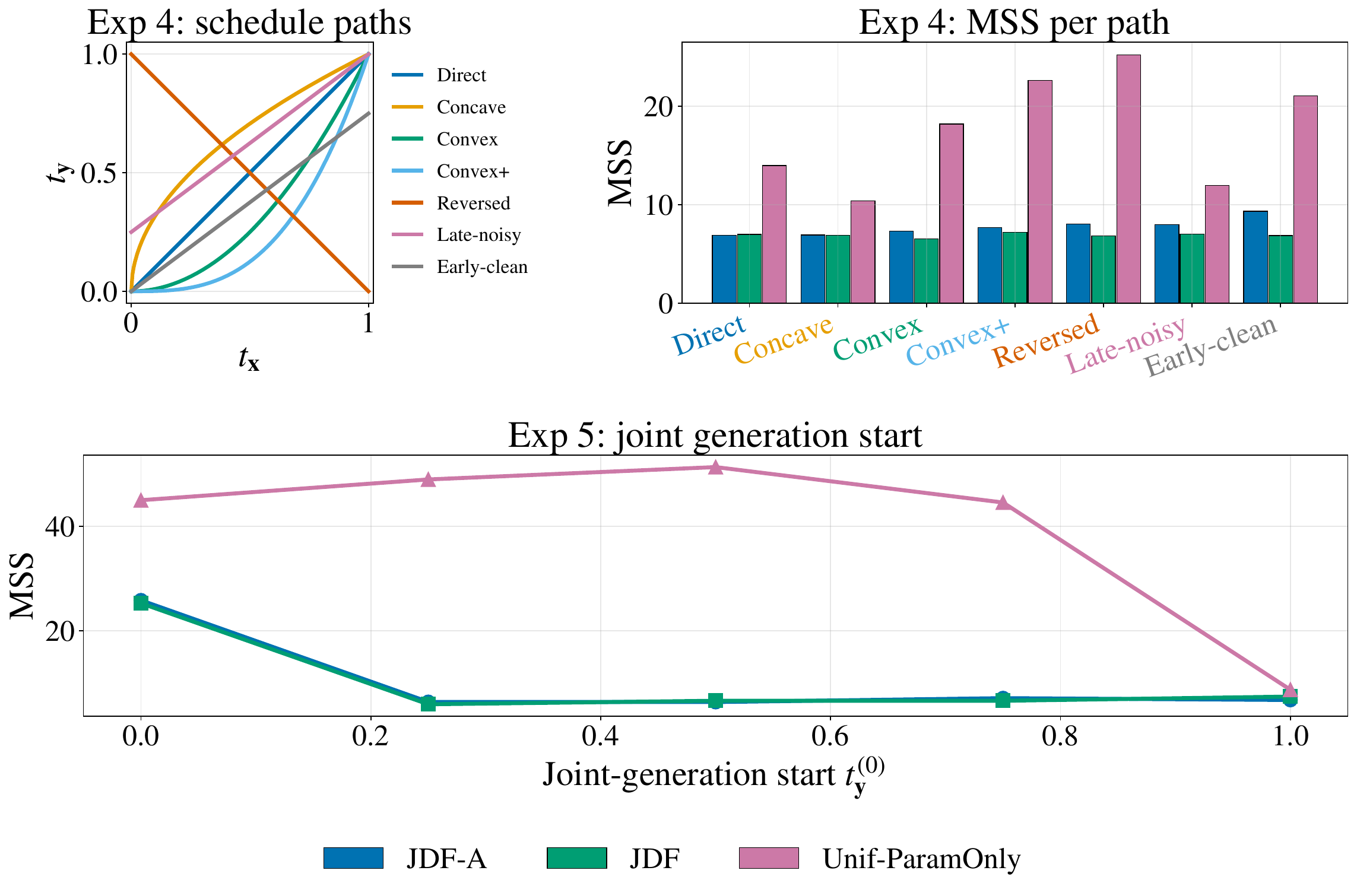}
  \caption{{Top left:} Schedules for time-varying conditioning noise. {Top right:} MSS on real audio for each schedule.  {Bottom:} joint sampling from a partially noised reference at $(0, t_\mathbf{y}^{(0)})$, integrating to $(1,1)$.}
  \label{fig:joint}
\end{figure}

Tables~\ref{tab:results-surge}~and~\ref{tab:results-dexed} report the main comparison. Both JDF and JDF-A substantially improve inversion of real audio compared to the Conditional baseline: on {\sc Surge XT}, MSS falls from $15.50$ to $7.68$, and on {\sc Dexed} from $13.03$ to $7.98$.
The same pattern holds across most remaining metrics, with the exception of SOT on {\sc Surge XT}.
Importantly, the off-manifold improvement is not achieved at the expense of \textit{in-distribution} performance: JDF(-A) also outperforms the baseline on synthetic audio across all metrics and both synthesizers.

While the Conditional baseline degrades sharply off-manifold (Surge {\sc XT} MSS $10.38\!\to\!15.50$), JDF degrades only marginally ($6.30\!\to\!7.68$).
This is likely explained, in part, by differences in the audio distributions.
The strong presence of one-shot audio samples for music production in the off-manifold dataset, for example, means the ends of many windows are silent.
In this sense, the framewise energy invariant SOT metric acts as an anchor, and indeed reveals that even for JDF-A, matching fine spectral details is more challenging off manifold.

\subsection{Training on off-manifold audio}
\label{subsec:realaudio}

To isolate the effect of the unpaired-real-audio marginal, we compare JDF-A to JDF, which shares the joint formulation but sees only synthesizer audio during training.
While we hypothesised that exposure to real recordings would further improve off-manifold reconstruction, JDF-A performs almost identically to JDF.
Off-manifold on {\sc Surge XT} the two are tied (MSS $7.
68$ vs.\ $7.69$), with JDF-A marginally ahead on other metrics. JDF performs slightly better in-distribution.
Nonetheless, this approach will allow future experimentation that incorporates auxiliary real audio such as stronger marginal weightings, generalisation to specialised off-manifold distributions, and representation-alignment objectives~\cite{yu_representation_2024,wu_representation_2025}.

\subsection{Isolating the effect of joint distribution training}
\label{subsec:cfmunified}

The JDF variants differ from the Conditional baseline both in objective and in conditioning architecture.
Conditional uses an AST audio encoder whose representation conditions a parameter DiT decoder, whereas JDF processes audio and parameter tokens jointly in a single shared DiT.
This removes the AST bottleneck and AdaLN-style conditioning~\cite{peebles_scalable_2023}, replacing them with direct self-attention between audio and parameter tokens.
Thus, any improvement over Conditional could reflect a more expressive conditioning pathway rather than the joint-density objective itself.
Unif-{\sc ParamOnly} controls for this possibility by using the same shared-token DiT as JDF while training only on the conditional inversion edge $\{t_\mathbf{y}=1\}$. 
Its slight improvement over the Conditional baseline suggests that the architecture makes a small contribution to the overall improvement.
However, its off-manifold performance remains worse than JDF and JDF-A.

This supports the central role of the joint distribution formulation.
When $\mathrm{supp}(\mu)=[0,1]^2$, most audio states encountered during training are partially noised rather than clean synthesized spectrograms.
As a result, the model learns a velocity field in a neighbourhood around the synthesizer manifold, making off-manifold inputs less of an extrapolation problem at inference.

\subsection{Classifier-free guidance for free}
\label{subsec:cfg}

JDF variants obtain an unconditional branch directly from the bottom edge $\{t_\mathbf{y}=0\}$, whereas Unif-{\sc ParamOnly} and the Conditional baseline require conditioning dropout to approximate the same branch. Fig.~\ref{fig:guidance} shows that CFG modestly improves in-domain performance for all models. Off manifold, however, CFG does not help the Conditional baseline: on {\sc Surge XT} its error grows monotonically with guidance strength (best at the lowest weight), and on {\sc Dexed} it is almost flat, with a shallow optimum near a weight of $3$.

We interpret this as a sign that CFG amplifies the Conditional model's already biased off-manifold predictions~\cite{ho_classifier-free_2021}.
By contrast, all other models exhibit improved performance under CFG, suggesting that this behaviour relates to the architectural differences identified in Section~\ref{subsec:cfmunified}.

\subsection{Noisy conditioning as posterior smoothing}
\label{subsec:noisy_cond}

Because $v_\theta$ is trained over the full time square, JDF models learn audio-conditioned parameter posteriors at intermediate audio noise levels, not only at $t_\mathbf{y}=1$. We can therefore condition on a partially noised reference,
$Y_\tau=(1-t_\mathbf{y})Y_0+t_\mathbf{y}\mathbf{y}^*$,
and sample from a smoothed posterior that interpolates between the parameter marginal at $\tau=0$ and the clean-conditioned posterior at $\tau=1$. Intuitively, lowering $\tau$ suppresses recording-specific detail that the synthesizer cannot reproduce while retaining coarse temporal and spectral structure.

Fig.~\ref{fig:noise_sweep} supports this view.
On-manifold, conditioning noise harms Conditional and Unif-{\sc ParamOnly}, which were trained for clean conditioning, while JDF variants remain stable except at the highest noise levels. 
Off-manifold, JDF and JDF-A suffer a slight degradation at low noise levels, but then improve sharply to reach their best reconstruction near $\tau=0.1$.
Unif-{\sc ParamOnly} still degrades under noise, but the Conditional baseline remarkably improves slightly off manifold under input noising.
This suggests that the Conditional baseline is highly sensitive to fine input detail under train-test mismatch, and benefits from smoothing, further motivating explicit training under these conditions.

We also vary the full path through the time square in Fig.~\ref{fig:joint}.
JDF and JDF-A are relatively insensitive to path choice, even under the pathological reversed path.
Unif-{\sc ParamOnly}, by contrast, performs poorly away from its training edge.
This suggests that the learned JDF velocity field remains useful across multiple regions of the time square, allowing inference-time path choices to affect reconstruction performance without retraining.

\subsection{Joint sampling from a partially noised reference}
\label{subsec:joint_sampling}
In the idealised flow-matching setting, paths within $\mathrm{supp}(\mu)$ correspond to transports between intermediate distributions. We therefore test whether such paths remain useful in the learned model.
Fig.~\ref{fig:joint} (bottom) integrates along a path beginning at $(0, t_\mathbf{y}^{(0)})$ and ending at $(1,1)$, allowing parameters and audio to co-evolve from a partially noised reference.
At $t_\mathbf{y}^{(0)}=1$ the path collapses to standard inversion; at $t_\mathbf{y}^{(0)}=0$ we sample the joint distribution unconditionally.
JDF and JDF-A are robust to the start level: error drops sharply once $t_\mathbf{y}^{(0)}\!\gtrsim\!0.25$ and stays flat up to the standard-inversion endpoint.
Unif-{\sc ParamOnly} fails everywhere except the degenerate $t_\mathbf{y}^{(0)}=1$ endpoint, again as expected: it has no training experience anywhere off the inversion edge.
Joint sampling is therefore possible in our trained models, but does not outperform fixed-noise conditioning in this setting. 
We report it primarily to illustrate that JDF models remain usable away from the conditional edge.

\section{Conclusion}\label{sec:conclusion}

We have shown that reframing synthesizer inversion as joint modelling of parameters and audio, rather than purely conditional estimation, improves robustness to real-world off-manifold inputs while preserving in-domain performance. By training a single multi-modal rectified flow over the time square $[0,1]^2$, paired synthesizer data and unpaired real recordings can be incorporated in one objective, and partially noised conditioning provides a simple inference-time mechanism for trading specificity against robustness. Experiments on {\sc Surge XT} and {\sc Dexed} show substantial improvements over a generative baseline on real-audio inversion, without sacrificing synthetic-audio performance.

Several directions remain open.
The flexibility of the joint distribution flow matching framework enables us to include further ``views'' of our training data which may further improve robustness, and unlocks further options for conditioning and control.
Moreover, our model's probabilistic framing suits it well to fine-tuning by reinforcement learning, which may serve to further refine the learnt distribution for off-manifold inputs.

We acknowledge some limitations of our presented approach.
First, noting that for off-manifold audio the notion of a matching synthesizer patch is loosely defined, we suggest that human evaluation of such a system should be a crucial component of subsequent work.
This should, secondly, be paired with an analysis of failure modes, as aggregate reconstruction metrics obscure potential commonalities between inputs which lead to worse outputs. 
Finally, certain design choices were selected heuristically and still warrant ablation.
Noting the significant effect that tuning the noise schedule has had on other diffusion and flow matching models~\cite{esser_scaling_2024}, for example, it is likely that more rigorous exploration of the time square sampling distribution will prove a fruitful avenue for future work.

\section{Acknowledgments}
With special thanks to Stefan Lattner and Tancr{\`e}de Martinez for the insightful discussions, and the entire Sony CSL Paris Music Team for their input and support.

\section{AI Usage Statement}

An AI coding assistant was used during the development of the research codebase, for writing experiment management tooling, and for collating and plotting results.
All AI code outputs were subject to extensive human verification to ensure the expected functionality was implemented.
The manuscript was drafted and edited with the aid of an LLM. 
However, all scientific claims, experimental results, and final wording were reviewed and approved by the author.

\section{Ethics Statement}

As in the work we build on~\cite{hayes_audio_2025}, we train exclusively on
synthetically generated data, publicly available audio datasets, and legally obtained private datasets.
Our choice of synthesizer reflects a bias toward the conventions of
western popular music production. 
While we expect our method to generalise to other synthesis paradigms, we have not empirically verified this. 
We view synthesizer inversion as a tool to augment, rather than replace, creative workflows.


\bibliography{references_camera_ready}


\end{document}

%% file: figures/timesquare.tex
\colorlet{joint}{blue!70!black}
\colorlet{jointbg}{blue!7}
\colorlet{condx}{red!75!black}
\colorlet{condy}{green!45!black}
\colorlet{margx}{orange!85!black}
\colorlet{margy}{pink!85!black}
\colorlet{noisycond}{yellow!60!black}

\begin{figure}[t!]
    \centering
    \resizebox{\linewidth}{!}{%
        \begin{tikzpicture}[
            >=Stealth,
            axis/.style = {->, thick},
            edge/.style = {line width=2.5pt, line cap=round},
            traj/.style = {->, line width=1.8pt, line cap=round},
            trajdash/.style = {line width=1.4pt, line cap=round, dashed},
            sample/.style = {circle, fill, inner sep=0pt, minimum size=2.6pt},
            edgesample/.style = {circle, fill, draw=white, line width=0.6pt,
                                 inner sep=0pt, minimum size=4pt},
        ]

        \begin{scope}[scale=7, local bounding box=left]

            \fill[jointbg] (0,0) rectangle (1,1);

            \draw[axis] (-0.04, 0) -- (1.12, 0) node[below, font=\large] {$t_\mathbf{x}$};
            \draw[axis] (0, -0.04) -- (0, 1.12) node[left, font=\large]  {$t_\mathbf{y}$};

            \draw[edge, condy]    (1,0) -- (1,1);   
            \draw[edge, condx]  (0,1) -- (1,1);   
            \draw[edge, margy] (0,0) -- (0,1);   
            \draw[edge, margx]   (0,0) -- (1,0);   

            \foreach \x/\y in {
                0.12/0.83, 0.22/0.41, 0.31/0.65, 0.43/0.18, 0.55/0.54,
                0.67/0.27, 0.78/0.72, 0.88/0.38, 0.17/0.23, 0.38/0.87,
                0.62/0.82, 0.83/0.55, 0.47/0.72, 0.72/0.12, 0.28/0.55,
                0.58/0.38, 0.19/0.68, 0.81/0.86, 0.36/0.30, 0.70/0.62,
                0.48/0.47, 0.25/0.92, 0.89/0.19, 0.09/0.47, 0.52/0.08}
                \node[sample, joint, opacity=0.75] at (\x,\y) {};

            \foreach \y in {0.08,0.28,0.47,0.62,0.78,0.92}
                \node[edgesample, condy]    at (1,\y) {};
            \foreach \x in {0.10,0.27,0.44,0.60,0.74,0.90}
                \node[edgesample, condx]  at (\x,1) {};
            \foreach \y in {0.08,0.25,0.42,0.58,0.75,0.91}
                \node[edgesample, margy] at (0,\y) {};
            \foreach \x in {0.09,0.26,0.43,0.59,0.76,0.93}
                \node[edgesample, margx]   at (\x,0) {};

            \draw (1,0) -- ++(0,-0.015) node[below] {$1$};
            \draw (0,1) -- ++(-0.015,0) node[left]  {$1$};
            \node[below left=1pt] at (0,0) {$0$};

            \node[font=\Large, joint, fill=blue!7, inner sep=2pt]
                at (0.5, 0.5) {$p(\mathbf{x},\,\mathbf{y})$};
            \node[condy,    font=\Large, xshift=14pt, rotate=-90]
                at (1, 0.5)  {$p(\mathbf{y}\mid\mathbf{x})$};
            \node[condx,  font=\Large, above=6pt]
                at (0.5, 1)  {$p(\mathbf{x}\mid\mathbf{y})$};
            \node[margy, font=\Large, xshift=-14pt, rotate=90]
                at (0, 0.5)  {$p(\mathbf{y})$};
            \node[margx,   font=\Large, below=6pt]
                at (0.5, 0)  {$p(\mathbf{x})$};

            \node[font=\Large\bfseries] at (0.5, 1.22) {Training};

        \end{scope}

        \begin{scope}[xshift=9.6cm, scale=7, local bounding box=right]

            \draw[thick, gray!55] (0,0) rectangle (1,1);

            \draw[axis] (-0.04, 0) -- (1.12, 0) node[below, font=\large] {$t_\mathbf{x}$};
            \draw[axis] (0, -0.04) -- (0, 1.12) node[left, font=\large]  {$t_\mathbf{y}$};

            \fill (0,0) circle (0.012);
            \fill (1,1) circle (0.012);
            \node[font=\scriptsize, below=12pt, xshift=6pt] at (0,0) {pure noise};
            \node[font=\scriptsize, above=8pt,  xshift=6pt] at (1,1) {clean data};

            \draw[traj, margx]   (0.00, 0) -- (1.0, 0);   
            \draw[traj, margy] (0, 0.00) -- (0, 1.0);   
            \draw[traj, condx]  (0.00, 1) -- (1.0, 1);   
            \draw[traj, condy]    (1, 0.00) -- (1, 1.0);   

            \draw[traj, noisycond] (0.01, 0.55) -- (0.99, 0.55);
            \draw[trajdash, noisycond, opacity=0.6] (0.01, 0.65) -- (0.99, 0.65);
            \draw[trajdash, noisycond, opacity=0.6] (0.01, 0.75) -- (0.99, 0.75);

            \draw[traj, joint] (0.02, 0.02) -- (0.97, 0.97);
            \draw[trajdash, joint, opacity=0.55]
                (0.02, 0.02) .. controls (0.25, 0.35) and (0.65, 0.85) .. (0.95, 0.95);
            \draw[trajdash, joint, opacity=0.55]
                (0.015, 0.02) .. controls (0.15, 0.48) and (0.4, 0.86) .. (0.95, 0.95);

            \draw (1,0) -- ++(0,-0.015) node[below] {$1$};
            \draw (0,1) -- ++(-0.015,0) node[left]  {$1$};
            \node[below left=1pt] at (0,0) {$0$};

            \node[condx,  font=\large, above=2pt]
                at (0.5, 1)  {audio-to-param};
            \node[condy,    font=\large, xshift=12pt, rotate=90]
                at (1, 0.5)  {param-to-audio};
            \node[margy, font=\large, xshift=-12pt, rotate=90]
                at (0, 0.5)  {uncond.\ audio};
            \node[margx,   font=\large, below=2pt]
                at (0.5, 0)  {uncond.\ params};

            \node[noisycond, font=\large, fill=white, inner sep=1.2pt,
                  anchor=west]
                at (0.65, 0.5) {noisy-cond.};
            \node[joint,   font=\large, rotate=45, fill=white, inner sep=1.2pt]
                at (0.28, 0.21) {joint};

            \node[font=\Large\bfseries] at (0.5, 1.22) {Inference};

        \end{scope}

        \end{tikzpicture}%
    }

    \caption{Flow-matching time $\mathbf{t}=(t_\mathbf{x},t_\mathbf{y})$ forms a
        unit square. \textbf{Training:} interior samples fit the joint
        density $p(\mathbf{x},\mathbf{y})$; samples on the top and right edges
        fit the conditionals $p(\mathbf{x}\mid\mathbf{y})$ and
        $p(\mathbf{y}\mid\mathbf{x})$; samples on the left and bottom edges fit
        the marginals $p(\mathbf{y})$ and $p(\mathbf{x})$.
        \textbf{Inference:} different sampling trajectories correspond to different tasks.}
    \label{fig:flow-square}
\end{figure}

%% file: tables/main.tex
\newcommand{\modelcol}[1]{\makebox[2.4cm][l]{#1}}

  \begin{table}[t]
  \centering
  \resizebox{\columnwidth}{!}{
  \begin{tabular}{l cccc@{}}
    \toprule
    & \multicolumn{4}{@{}c}{{\sc Surge XT}} \\
    \cmidrule(lr){2-5}
    \modelcol{Model} & MSS $\downarrow$ & wMFCC $\downarrow$ & SOT $\downarrow$ & RMS $\uparrow$ \\
    \midrule
    \multicolumn{5}{@{}l}{\textit{(a) On-manifold} (Synth audio)} \\
    \modelcol{JDF}                 & \textbf{6.30} & \textbf{7.96} & \textbf{0.063} & \textbf{0.940} \\
    \modelcol{JDF-A}               & 6.61 & 8.57 & 0.063 & 0.939 \\
    \modelcol{Unif-{\sc ParamOnly}} & 9.13 & 11.41 & 0.096 & 0.910 \\
    \modelcol{Conditional}         & 10.38 & 12.73 & 0.111 & 0.907 \\
    \addlinespace
    \multicolumn{5}{@{}l}{\textit{(b) Off-manifold} (Real audio)} \\
    \modelcol{JDF}                 & \textbf{7.68} & 9.37 & 0.317 & 0.781 \\
    \modelcol{JDF-A}               & 7.69 & \textbf{9.34} & 0.309 & \textbf{0.787} \\
    \modelcol{Unif-{\sc ParamOnly}} & 11.35 & 12.76 & 0.282 & 0.743 \\
    \modelcol{Conditional}         & 15.50 & 14.91 & \textbf{0.232} & 0.700 \\
    \bottomrule
  \end{tabular}
  }
  \caption{Audio reconstruction results on {\sc Surge XT}.}
  \label{tab:results-surge}
  \end{table}

  \begin{table}[t]
  \centering
  \resizebox{\columnwidth}{!}{
  \begin{tabular}{l cccc@{}}
    \toprule
    & \multicolumn{4}{@{}c}{{\sc Dexed}} \\
    \cmidrule(lr){2-5}
    \modelcol{Model} & MSS $\downarrow$ & wMFCC $\downarrow$ & SOT $\downarrow$ & RMS $\uparrow$ \\
    \midrule
    \multicolumn{5}{@{}l}{\textit{(a) On-manifold} (Synth audio)} \\
    \modelcol{JDF-A}       & \textbf{7.28} & \textbf{13.16} & \textbf{0.042} & \textbf{0.916} \\
    \modelcol{Conditional} & 10.30 & 16.79 & 0.060 & 0.890 \\
    \addlinespace
    \multicolumn{5}{@{}l}{\textit{(b) Off-manifold} (Real audio)} \\
    \modelcol{JDF-A}       & \textbf{7.98} & \textbf{15.25} & \textbf{0.511} & \textbf{0.816} \\
    \modelcol{Conditional} & 13.03 & 20.66 & 0.553 & 0.762 \\
    \bottomrule
  \end{tabular}
  }
  \caption{Audio reconstruction results on {\sc Dexed}.}
  \label{tab:results-dexed}
  \end{table}